\documentclass[pre,aps,twocolumn,superscriptaddress,amsmath,showpacs,floatfix]{revtex4}
\usepackage{graphicx}
\begin{document}
\title{Entropic attraction of adhesion bonds toward cell boundaries}
\author{Noam Weil and Oded Farago}
\affiliation{Department of Biomedical Engineering, Ben Gurion University,
Be'er Sheva 84105, Israel}
\begin{abstract}
Adhesion bonds between membranes and surfaces are attracted to each
other via effective interactions whose origin the entropy loss due to
the reduction in the amplitude of the membrane thermal fluctuations in
the vicinity of the adhesion bonds. These fluctuation-induced
interactions are also expected to drive the adhesion bonds toward the
rim of the cell, as well as toward the surfaces of membrane
inclusions. In this paper, we analyze the attraction of adhesion bonds
to the cell inner and outer boundaries. Our analysis shows that the
probability distribution function of a single (diffusing) adhesion
bond decay algebraically with the distance from the boundaries. Upon
increasing the concentration of the adhesion bonds, the attraction to
the boundaries becomes strongly self-screened.
\end{abstract}
\maketitle

\section{Introduction}
\label{sec:1}

Biological membranes are essential to every living organism. They
consist of a bilayer of amphiphilic lipids, proteins and cholesterol,
and serve as a barrier between the cytosol and the extracellular
matrix in biological cells \cite{Alberts:1994}. In eukaryotic cells,
membranes surround the organelles, allowing for organization of
biological processes through compartmentalization. Membranes are
involved in a variety of cellular processes, including the transition
of nutrients in and out of the cell, signal transduction and
communication \cite{Beckerle:2001}.

One of the processes taking place at the cell membrane is cell
adhesion. This process, during which the cell membrane is attracted to
another interface (which may be the membrane of another cell) can, in
principle, be facilitated by non-specific attractive interactions
(e.g., Coulomb and van der Waals interactions). Most often, however,
cell adhesion is mediated by specific binding between receptors that
reside on the membrane surface and ligands located on the opposite
surface \cite{Linderman:1995}. Specific adhesion usually occurs in
regions with high density of receptors and ligands. When facing a
surface with enough ligands, the receptors may cluster into highly
concentrated adhesion domains to establish strong binding
\cite{Smith:2007,Lipowsky:2009}. Specific bioadhesion occurs in a
variety of cellular processes, including binding of white blood cells
to pathogens \cite{Naggli:2003}, binding and fusion of drug carrier
liposome to target cells \cite{Cullis:2004}, cadherin-mediated
adhesion of neighboring cells \cite{Giehl:2008}, focal adhesion of
cells to the extracellular matrix \cite{Yamada:2001}, and cell
signaling.

Adhesion induced domain formation results from both the activity of
the actin cytoskeleton, and from attractive interactions which are
mediated by the membrane curvature elasticity \cite{riveline:2004}.
The membrane-mediated interactions give rise to two distinct, but not
entirely unrelated, mechanisms facilitating cluster formation. The
first mechanism is the Casimir-like attraction between two bonds,
which originates from the reduction in the amplitude of the membrane
thermal fluctuations in the vicinity of the bonds and the associated
entropy loss. The resulting increase in the free energy of the
membrane is minimized when the two bonds come within close proximity
of each other, in which case the area where the thermal fluctuations
are suppressed is reduced \cite{bruinsma:1994}. The second mechanism
is related to the tendency of new bonds to form next to existing
ones. This ``binding cooperativity'' effect stems from the fact that
the ligand-receptor binding probability is highest when the separation
between the ligand and the receptor on the opposite surfaces is close
to the bond length - conditions which are likely to occur near already
existing bonds \cite{reister:2011}. In addition, since the formation
of new bonds leads to an overall reduction in the membrane roughness,
it also increases the binding probability and assists the formation of
even more bonds \cite{weikl:2009}. Generally speaking, the first
mechanism is likely to contribute to the aggregation of irreversible
bonds with large binding energies, while the second one is probably
more relevant to the formation of metastable domains of weaker
reversible bonds.

In a previous publication we proposed a framework for studying the
Casimir-like interactions between adhesion bonds \cite{Weil:2010}. Our
analysis is based on the following idea: The Casimir-like interactions
are induced by the membrane thermal fluctuations whose amplitude is
suppressed due to the presence of adhesion bonds. To calculate the
free energy cost of a given distribution of adhesion bonds, a free
energy density is assigned to each unit area of the fluctuating
membrane, and the free energy density is integrated over the entire
membrane area. The free energy density reflects the extend by which
the amplitude of the fluctuations is locally reduced compared to the
fluctuations of a free (non-adhering) membrane. The key point of the
method is the assumption amplitude of thermal fluctuations depends on
the distance $d_{\rm min}$ to the nearest adhesion bond. Accordingly,
the free energy density associated with each unit area of the membrane
is given by (see derivation in \cite{Weil:2010}):
\begin{equation}
V(r)=\left(\frac{k_BT}{\pi}\right)\left(\frac{1}{d_{\rm min}}\right)^2.
\label{eq:2}
\end{equation} 
The total attachment free energy is given by
\begin{equation}
F=\int V\left(d_{\rm min}\left(\vec{r'}\right)\right)d\vec{r'}=
\sum_{i}\frac{k_BT}{\pi}\left(\frac{l}{d_{\rm min}}\right)^2\left(
1-s_i\right),
\label{eq:3}
\end{equation}
where the integration is carried over the entire membrane area accept
for the immediate vicinities of the adhesion bonds. The sum on the
r.h.s.~of eq.~\ref{eq:3} is the discrete analog of the integral, which
is useful for lattice simulations and numerical integrations. In the
sum, the variable $s_i$ takes the value $s_i=1$ for the sites occupied
by adhesion bonds and $s_i=0$ for an empty sites that represent
fluctuating unit areas of size $l^2$. Only the latter contribute to
the sum.

We studied the model defined by eq.~\ref{eq:3} by using both a mean
field analysis and Monte Carlo simulations \cite{Weil:2010}. Our
analysis revealed (in agreement with other theoretical studies
\cite{weikl:2007,speck:2011}) that the fluctuation-induced
interactions are not strong enough to allow the formation of adhesion
domains. However, usually the adhesion bonds also interact with each
other via short range attractive hydrophobic and depletion forces
\cite{Israelachvili:1985}; and when these residual interactions are
introduced into the model, condensation of adhesion domains becomes
thermodynamically possible \cite{Weil:2010,weikl:2007}. Our analysis
showed that the fluctuation-induced Casimir-like interactions greatly
reduce the strength of the residual attractive interactions required
for cluster formation by almost half an order of magnitude, to below
the thermal energy $k_BT$.

In this work we take a step forward and study the formation of
adhesion domains in cellular membranes which also include
trans-membrane proteins. Since these inclusions significantly
influence the membrane spectrum of thermal fluctuations
\cite{Farago:2007}, their presence is expected to modify the
fluctuation-induced interaction between adhesion bonds and their
aggregation behavior. In this context, it is worthwhile to mention
that there is a great bulk of experimental
\cite{Boxer:1996,Groves:2006,Urbach:2002} and theoretical
\cite{Pincus:1996,Lubensky:1996,Golestanian:1999,Oster:2000} works on
the fluctuation-mediated potential between the inclusions
themselves. Here, we tackle a different problem and explore how a
single inclusion affects the aggregation of adhesion bonds. Since the
formation of adhesion cluster is driven by the tendency to localize
the reduction of the thermal fluctuations in a restricted area, it is
natural to expect that adhesion bonds would exhibit affinity to the
surface of the inclusion. Physically, the effect of an inclusion on
the thermal fluctuations should be similar to that of a cluster of
adhesion bonds of the same size. However, as discussed above, without
direct residual interactions the system is in the gas-phase and a
cluster of adhesion bonds is not stable. Therefore, the presence of
the inclusion is not equivalent to the formation of a cluster, which
is only transient.

The paper is organized as follows: In sec.~\ref{sec:2}, we examine the
attraction between a circular inclusion and a single adhesion bond. We
derive the probability distribution of the adhesion bond from the
center of the inclusion. We show that in contrast to the weak
attraction between two adhesion bonds (see ref.~\cite{Farago:2010}),
the fluctuation-induced attraction between the adhesion bond and a
large inclusion is sufficiently strong to keep the adhesion bond in
the vicinity of the inclusion. In sec.~\ref{sec:3}, we study the
distribution and aggregation of many adhesion bonds in the presence of
the inclusion. Using Monte Carlo simulations with different boundary
conditions and different adhesion bonds densities, we demonstrate that
the strong attractive potential calculated in sec.~\ref{sec:2} is
screened and felt only by the adhesion bonds located near the surface of
the inclusion. In sec.~\ref{sec:4}, we summarize and discuss our
results.

\section{The bond-inclusion pair interaction}
\label{sec:2}

\begin{figure}[t]
\begin{center}
\scalebox{0.45}{\centering \includegraphics{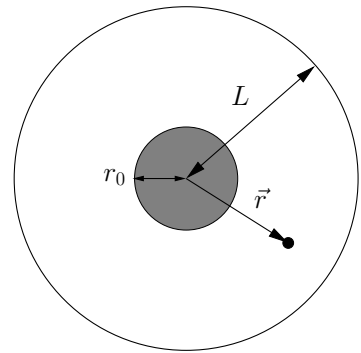}}
\end{center}
\vspace{-0.5cm}
\caption{Schematic of the system under investigation, consisting of a
circular membrane of radius $L$ with a circular inclusion of radius
$r_0$ at the center. The membrane is supported by a flat surface to
which it is attached by a single adhesion bond located at $\vec{r}$.}
\label{fig:1}
\end{figure}

We consider the model system shown schematically in fig.~\ref{fig:1},
consisting of a circular membrane of radius $L$, a circular inclusion
of radius $r_0\ll L$ fixed at the center of the membrane, and a single
adhesion bond located at $\vec{r}$. The total attachment free energy
of the system is given by eq.~\ref{eq:3}
\begin{equation}
F(\vec{r},r_0,L)=\int_{r_0+l}^{L-l}\frac{k_BT}{\pi d_{\rm min}^2
(\vec{r'},\vec{r},r_0,L)}d\vec{r'},
\label{eq:4}
\end{equation} 
where the integration is carried over the whole area of the membrane,
excluding regions of microscopic size $l$ near the inner
($|\vec{r'}|=r_0$) and outer ($|\vec{r'}|=L$) boundaries, and in the
vicinity of the adhesion bond ($\vec{r'}=\vec{r}$). In eq.~\ref{eq:4},
$d_{\rm min}$ denotes the distance of a given point $\vec{r'}$ on the
membrane to the nearest ``obstacle'', which may be either the adhesion
bond or the surface of the inclusion. The assumption underlying
eq.~\ref{eq:4} is that the membrane separation at the inclusion is the
same as the length of the adhesion bond. This feature is also likely
to contribute to the binding cooperativity mechanism mentioned above
in section \ref{sec:1}, which we do not consider in this work.  Had
there been a length mismatch between the inclusion and the adhesion
bonds, it would have involved a bending elasticity energy cost that
would weaken the fluctuation-induced attraction.  In a binary mixtures
of adhesion bonds, the inclusion is likely to have a higher affinity
to the adhesion bonds with the smaller length mismatch, which may lead
to phase segregation in the mixture \cite{weikl:2010,burroughs:2011}.

In what follows, we consider membranes with both ``open'' and
``closed'' outer boundaries. At a closed boundary, the membrane is
attached to a frame, whereas at an open boundary, the membrane
fluctuates freely. Thus, the former is also considered as an obstacle,
while the latter is not. The open and closed boundary conditions
represent two limiting cases. Cellular membranes are attached to the
actin cytoskeleton located mainly at the periphery of the cell., which
corresponds to an intermediate case between open and closed boundary
conditions.

\begin{figure}[t]
\begin{center}
\scalebox{0.6}{\centering \includegraphics{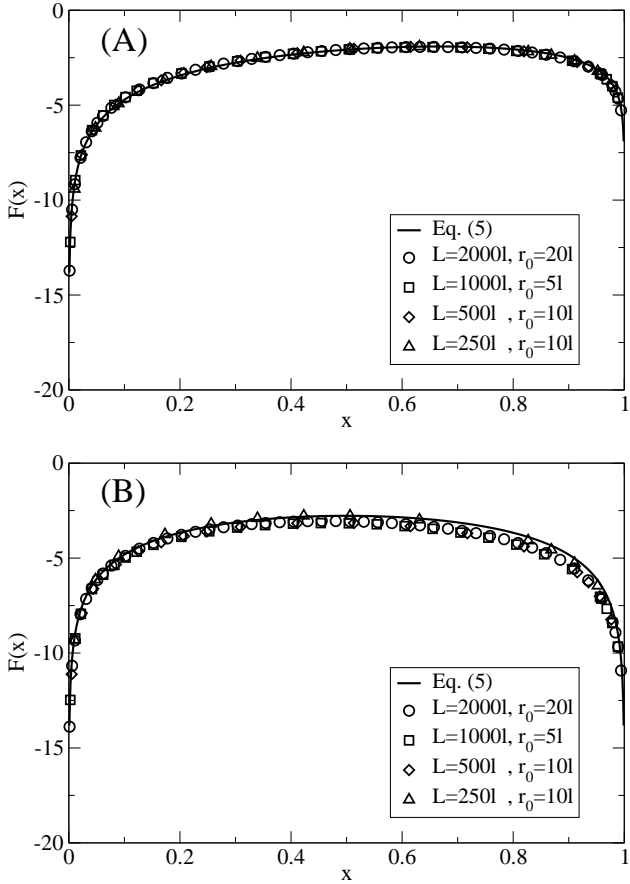}}
\end{center}
\vspace{-0.5cm}
\caption{Data collapse of the attachment free energy for different
values of $|\vec{r}|$, $r_0$, and $L$, as a function of the
dimensionless distance $x=(|\vec r|-r_0)/(L-r_0)$. The data in (A) and
(B) corresponds to systems with open and closed outer boundaries,
respectively.}
\label{fig:2}
\end{figure}

The integral in eq.~\ref{eq:4} has been evaluated numerically for
different values of $r_0$ and $L$, and for $r_0<|\vec{r}|<L$. Our
calculations reveal that the attachment free energy is very well
approximated by the following expression
\begin{eqnarray}
F(\vec r,r_0,L)&\sim&2k_BT\ln{\frac{(|\vec r|-r_0)}{l}} 
\nonumber \\
&+&nk_BT\ln{\frac{(L-|\vec r|)}{l}}+C(r_o,L),
\label{eq:5}
\end{eqnarray}
with $n=1$ for an open outer boundary, and $n=2$ for a closed
boundary. Introducing the dimensionless distance $0<x=(|\vec
r|-r_0)/(L-r_0)<1$, eq.~\ref{eq:5} can be also expressed as
\begin{eqnarray}
F(x)&\equiv& F(\vec
r,r_0,L)-\ln{\left(\frac{L-r_0}{l}\right)^{2+n}}
-C(r_o,L) \nonumber \\
&\sim&
2k_BT\ln(x)+nk_BT\ln(1-x).
\label{eq:6} 
\end{eqnarray}
Our numerical results along with eq.~\ref{eq:6} are plotted in
figs.~\ref{fig:2}(A) and \ref{fig:2}(B) for an open and closed boundary,
respectively. When the adhesion bond is located far away from the
outer boundary, i.e. for $x\ll 1$, the attachment free energy is
dominated by the first term $F\simeq 2k_BT\ln(x)$, which has been
previously derived as the pair potential of mean force between two
adhesion bonds \cite{Farago:2010}.  In the vicinity of an outer closed
boundary ($x\lesssim 1$), one gets $F\simeq 2k_BT\ln(1-x)$, which
implies that the adhesion bond is attracted to the nearest point on
the outer closed boundary as if this point is another adhesion
bond. For an open boundary, the prefactor of the second term in
eq.~\ref{eq:6} is reduced by half from $n=2$ to $n=1$. This can be
rationalized by noticing that an open boundary is identical to the
midplane between two adhesion bonds in a twice larger system. The pair
potential between the adhesion bond and its image is
$2k_BT\ln[2(1-x)]=2k_BT\ln[(1-x)]+C$, and this energy is equally
divided between the real and imaginary halves of the system.

\begin{figure}[t]
\begin{center}
\scalebox{0.35}{\centering \includegraphics{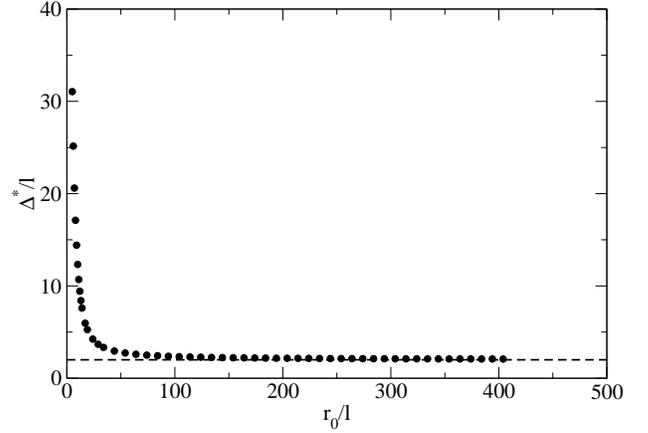}}
\end{center}
\vspace{-0.5cm}
\caption{The width $\Delta^*$ of the shell around the inclusion, as a
function of the radius of the inclusion $r_0$. The dashed horizontal
line corresponds to the solution of eq.~\ref{eq:10}, $\Delta^*=2l$,
for asymptotically large $r_0$.}
\label{fig:3}
\end{figure}

How strong is the attraction between the adhesion bond and the inner
inclusion? From eq.~\ref{eq:5} we find that the probability
distribution function $P(\vec{r})\sim
\exp{\left[-F(\vec{r},r_0,L)/k_BT\right]}$ of the adhesion bond is
given by
\begin{equation}
P(\vec{r})=\frac{1}{Z}\frac{1}{(|\vec{r}|-r_0)^2(L-|\vec{r}|)^n},
\label{eq:7}
\end{equation}
where the normalization factor is
\begin{equation}
Z=\int_{r_0+l}^{L-l}\frac{2\pi rdr}{(r-r_0)^2(L-r)^n}.
\label{eq:8}
\end{equation}

In the case of an open outer boundary ($n=1$), the mean distance
between the adhesion bond and the surface of the inner inclusion is
given by
\begin{equation}
\left\langle r \right\rangle -
r_0\equiv\Delta=\frac{(L^2-r_0^2)\ln{(\frac{L-r_0-l}{l})}}
{2L\ln{(\frac{L-r_0-l}{l})}+r_0(r_0-L)\left[\frac{2l+r_0-L}{l(L-r_0-l)}
\right]},
\label{eq:9}
\end{equation}
which for $L\gg r\equiv|\vec{r}|\gg r_0$, simplifies to $\Delta\sim
L/2$. The fact that the mean separation $\Delta$ grows linearly with
the size of the membrane may be considered as an indication that the
adhesion bond is not bound to the inclusion. However, a closer
inspection of eq.~\ref{eq:7} reveals that the adhesion bond spends
most of its time near the boundary of the inner inclusion. To quantify
this phenomenon, we introduce the length scale $\Delta^*$, which is the
width of the shell around the inner inclusion where the probability to
find the adhesion bond is $0.5$. The length $\Delta^*$ is determined
by solving the equation

\begin{widetext}
\begin{equation}
P_{\rm acc}(\Delta^*)=\int_{r_0+l}^{r_0+\Delta^*}2\pi|\vec{r}|P(|\vec{r}|)dr=
\frac{2L\ln{\left(\frac{\Delta^*}{l}\right)}+r_0(r_0-L)(\frac
{l-\Delta^*}{l\Delta^*})}{2L\ln{(\frac{L-r_0-l}{l})}+r_0(r_0-L)
\left[\frac{2l+r_0-L}{l(L-r_0-l)}\right]}=0.5.
\label{eq:10}
\end{equation}
\end{widetext}
Our results for $\Delta ^*$ are plotted in fig.~\ref{fig:3} as a
function of $r_0$ and for $L\gg r_0$. The results shows that
$\Delta^*$ is a monotonically decreasing function of $r_0$. For very
small inclusions of size $r_0\sim l$, the length $\Delta^*\gg l$. This
result is consistent with the observation of ref.~\cite{Farago:2010}
that the fluctuation-mediated pair potential is too weak to bind two
adhesion bonds to each other. For inclusions of size $r_0\gtrsim 20l$,
the length $\Delta^*<5l$; and for asymptotically large inclusions
($r_0\rightarrow \infty$), $\Delta^*\rightarrow 2l$. In other words,
although $\Delta\sim L/2$, the adhesion bond is likely to be found
within a thin shell around the surface of a sufficiently large
inclusion.

\begin{figure}[t]
\begin{center}
\scalebox{0.35}{\centering \includegraphics{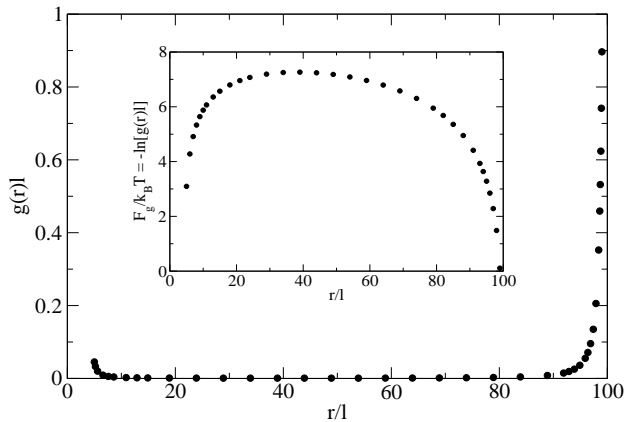}}
\end{center}
\vspace{-0.5cm}
\caption{The radial distribution function $g$ as a function of $r$,
for $r_0=4l$ and $L=100l$. The inset shows the corresponding free
energy, $F_g=-k_BT\ln[g(r)l]$.}
\label{fig:4}
\end{figure}

For a closed outer boundary, the radial distribution function of the 
adhesion bond
\begin{widetext}
\begin{equation}
g(r)\equiv 2\pi r P(|\vec{r}|)=\frac{r}{(r-r_0)^2(L-r)^2}
\left\{
\frac{(L-r_0)^3}{(L+r_0)\left[2\ln{\left
(\frac{L-r_0-l}{l}\right)}+(r_0-L)\left(\frac{2l+r_0-L}{l(L-r_0-l)}\right)
\right]}\right\}.
\label{eq:11}
\end{equation}
\end{widetext}
(The expression appearing in braces in eq.~\ref{eq:11} is a
normalization constant that depends on $r_0$ and $L$). The radial
distribution function $g(r)$, which is plotted in fig.~\ref{fig:4} for
$r_0=4l$ and $L=100L$, increases rapidly as one approaches both the
inner and outer boundaries of the membrane. However, unlike the
probability density per unit area $P(\vec{r})$ which is symmetric with
respect to the mid-radius $r_m=(L+r_0)/2$, the radial distribution
function $g(r)\sim rP(|\vec{r}|)$ is much higher near the outer
boundary than the inner one. This, of course, is directly related to
the fact that there is simply more membrane area near the outer
boundary. In the example plotted in fig.~\ref{fig:4} ($r_0=4l$,
$L=100l$), the probabilities to find the adhesion bond within a shell
of size $4l$ around the outer and inner boundaries are 72.5\% and
4.5\%, respectively. For a bigger membrane of size $L=1000l$ (which,
assuming that $l$ is of the order of a few nanometers, is a reasonable
estimate for the size of the cell plasma membrane), these
probabilities change to 79\% and 0.5\%, respectively. These numbers
suggest that the adhesion bond is likely to be ``adsorbed'' at the
outer boundary. This, however, is only part of the story. If the
adhesion bond is located near the inner boundary, it is not going to
easily escape to the outer rim. The inset on fig.~\ref{fig:4} shows
the potential of mean force $F_g(r)\equiv
-k_BT\ln\left[g\left(r\right)\right]$, exhibiting a free energy
barrier of $\sim 4k_BT$ that blocks the migration of the adhesion bond
from the inner boundary to the outer one. The barrier on the opposite
direction is of $\sim 7k_BT$. For $L=1000l$, the free energy barriers
increase to $\sim 7k_BT$ and $\sim 12k_BT$ for the escape from the
inner and outer boundaries, respectively.  Thus, as in the case with
an open boundary discussed above, an adhesion bond located near the
inner boundary may remain trapped there for relatively long times.

\section{Distribution of adhesion bonds around an inclusion}
\label{sec:3}

The previous section dealt with the localization of a single adhesion
bond near the boundaries of the system. Cellular adhesion, however,
involves an ensemble of adhesion bonds which may cluster into adhesion
domains. In a previous work \cite{Weil:2010}, we demonstrated that the
formation of adhesion domains cannot be induced by the Casimir-like
interactions alone, and that it requires the existence of residual
attractive interactions between the adhesion bonds. Here, we wish to
understand whether this picture is changed by the presence of a
membrane inclusion that serve as a potential nucleation seed for an
adhesion domain. One can envision a scenario where a thin shell of
adhesion bonds is formed around the inclusion, effectively increasing
its size and promoting the recruitment of more adhesion bonds. The
opposite scenario, in which the addition of adhesion bonds self-screen
the attraction toward the inclusion, is also plausible. In order to
investigate this issue, we performed Monte Carlo lattice simulations
of the model defined by eq.~\ref{eq:3} on circular systems with
$r_0=4l$ and $L=100l$ (where $l$ is the lattice spacing). Our
simulations were conducted at relatively low concentrations of
adhesion bonds, $\phi=0.01$ and $\phi=0.05$, with both open and closed
boundary conditions.

\begin{figure}[t]
\begin{center}
\scalebox{0.6}{\centering \includegraphics{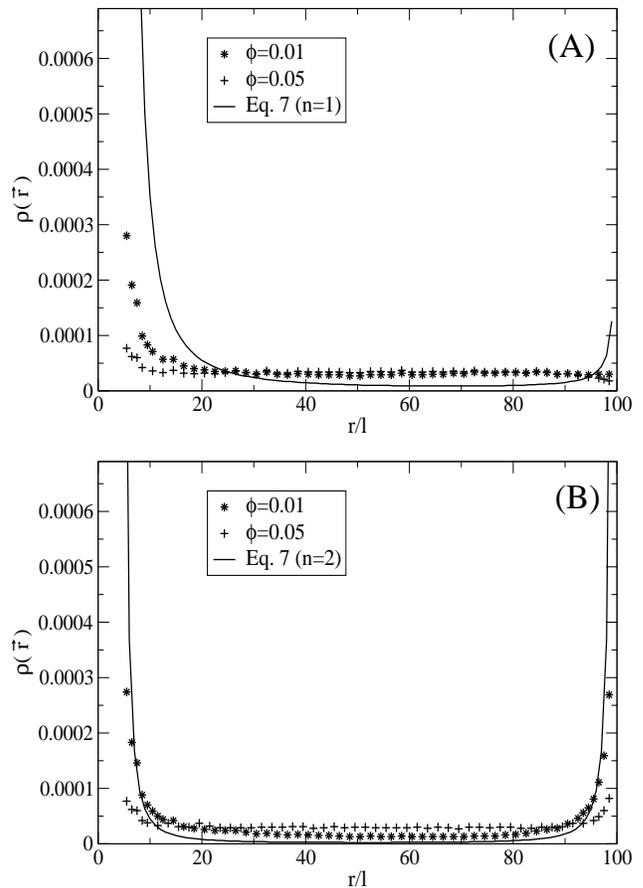}}
\end{center}
\vspace{-0.5cm}
\caption{The normalized number density $\rho(\vec{r})$ as a function
of $r=|\vec{r}|$. Data shown in the figure corresponds to $r_0=4l$ and
$L=100l$, for systems with (A) open and (B) closed outer
boundaries. The solid lines depict the probability distribution of a
single adhesion bond.}
\label{fig:5}
\end{figure}

We measured the number density (per unit area) of adhesion bonds as a
function of $\vec{r}$. To allow comparison with the probability
distribution $P(\vec{r})$ (eq.~\ref{eq:7}), we normalized the number
density to the total number of adhesion bonds. Our results for the
normalized number density, $\rho(\vec{r})$, are depicted in
fig.~\ref{fig:5}(A) and \ref{fig:5}(B) for open and closed boundaries,
respectively. Both figures demonstrate that the fluctuation-induced
attraction to the boundaries is self-screened by the presence of
adhesion bonds. Comparison of our results for $\phi=0.01$ (depicted by
stars in both figures) and $\phi=0.05$ (pluses), with eq.~\ref{eq:7}
for a single adhesion bond (solid lines), reveals that the
self-screening effect is increased with the increase in the
concentration $\phi$. For an open boundary, the slight increase in
$P(\vec{r})$ near the outer boundary disappears already at
$\phi=0.01$. For $\phi=0.05$, the density $\rho(\vec{r})$ is almost
constant with no significant affinity of adhesion bonds toward the
inner boundary. The same trends are also observed in the case of a
closed boundary, where $P(\vec{r})$ increases very sharply near both
boundaries. When the concentration $\phi$ increases, these maxima are
quickly lowered, and the number density $\rho(\vec{r})$ becomes almost
uniform.

\section{Summary and Discussion}
\label{sec:4}

In this work we continued our investigation of the entropic,
fluctuation-induced, attraction between adhesion bonds in supported
biological membranes. The focus here was on the possible role played
by the membrane boundaries as nucleation seeds for the formation of
adhesion domains. Residence of adhesion bonds near the cell boundaries
is thermodynamically favorable because it lowers the entropic cost
associated with the suppression of the membrane thermal fluctuations
around the adhesion bonds. We started our investigation by looking at
the pair interaction between an adhesion bond and a circular membrane
inclusion, and calculating the distribution function of the adhesion
bond around the inclusion. In the case of an open outer cell boundary,
the probability density of the adhesion bond decays algebraically with
the distance from the surface of the inclusion. Although the mean pair
distance grows linearly with the size of the membrane, the
distribution function is such that the adhesion bond spends most of
its time in the immediate vicinity of the inclusion. When the outer
membrane boundary is closed, the probability distribution per unit
area is symmetric with respect to the mid radius. In this case, the
adhesion bond is likely to be found near the outer boundary, which is
much larger than the inner one. Diffusion of the adhesion bonds
between the two boundaries is limited by the existence of a
substantial free energy barrier in the middle of the membrane.

Does this analysis imply that the cell boundaries can serve as
nucleation seeds for adhesion domains? Perhaps yes, but probably not
due to the fluctuation-induced mechanism alone. Our simulation results
show that even at small densities of adhesion bonds, the entropic
attraction toward the cell boundaries are entirely
self-screened. This observation is consistent with our previous study
of systems with periodic boundary conditions (no boundaries), where it
has been found that the fluctuation-induced interactions are not
sufficient to allow the formation of adhesion domains. In other words,
without additional attractive interactions, the problem of aggregation
of adhesion bonds is similar to the problem of gas to liquid
condensation {\em above}\/ the transition temperature. Introducing a
nucleation seed into such a system without changing the temperature is
not going to drive the system into the condensed phase. We leave for a
future work the problem whether domain formation occurs at high
density of inclusions whose inter-distances are smaller than the
screening length for the adhesion bonds. Another aspect of the problem
that needs to be explored is whether the existence of direct residual
attractive interaction between the adhesion bonds and the inclusion
can facilitate cluster formation, even in the absence of direct
attractive interactions between the adhesion bonds themselves.

We thank Haim Diamant and Thomas Weikl for helpful comments. This
work was supported by the Israel Science foundation (Grant Number
946/08).


\end{document}